\documentclass[a4paper,11pt]{article}
\usepackage{jinstpub} 
\usepackage{lineno}
\usepackage{gensymb}
\usepackage{lineno}
\usepackage{amsmath}
\usepackage{xspace}
\usepackage{graphicx}
\usepackage{hyperref}
\usepackage{url}

\title{\boldmath Performance characterisation of the Hamamatsu R760 photomultiplier tube for the PLUME detector}

\author[a,b,f,1]{A. Bellavista,\note{Corresponding author.}}
\author[b,f]{A. Carbone,}
\author[c]{V. Chaumat,}
\author[b,f]{F. Ferrari,}
\author[c]{T. Nguyen-Trung,}
\author[c]{V. Puill,}
\author[d]{L. Toscano,}
\author[e]{and A. Villa}

\affiliation[a]{European Organization for Nuclear Research (CERN), Geneva, Switzerland}
\affiliation[b]{Università di Bologna - Dipartimento di Fisica e Astronomia "Augusto Righi", Bologna, Italy}
\affiliation[c]{Laboratoire de Physique des 2 Infinis Irène Joliot-Curie, IJCLab, Orsay, France}
\affiliation[d]{Fakultät Physik, Technische Universität Dortmund, Dortmund, Germany}
\affiliation[e]{Institute of Physics, Ècole Polytechnique Fèdèrale de Lausanne (EPFL), Lausanne, Switzerland}
\affiliation[f]{Istituto Nazionale di Fisica Nucleare - Sezione di Bologna, Bologna, Italy}

\emailAdd{alberto.bellavista@cern.ch}

\abstract{The Probe for Luminosity Measurement detector is a novel luminometer designed to monitor the luminosity and beam conditions of the Large Hadron Collider at the interaction point of the LHCb experiment, starting from Run 3. The detector is based on a hodoscope composed of 48 Hamamatsu R760 photomultiplier tubes, which detect the Cherenkov light produced by charged particles originating from the interaction region. The accurate and stable operation of these sensors is essential to ensure reliable luminosity measurements throughout the full data-taking period.

This paper presents a detailed characterisation of the photomultiplier tubes currently installed in the detector. In particular, their absolute gain, transit-time drift, linearity, dark current, and ageing behaviour are systematically studied under controlled laboratory conditions. The results provide a comprehensive assessment of the performance of the detection modules and establish the optimal operating conditions required to ensure stable and precise measurements throughout Run 3 and Run 4.}


\begin{document}
\maketitle
\flushbottom

\section{Introduction}
\label{sec:intro}

The LHCb experiment is one of the nine particle physics detectors installed at the Large Hadron Collider (LHC) at CERN. Its primary research focus is flavour physics, with particular emphasis on the study of charge conjugation and parity (CP) violation in beauty and charm hadrons produced in proton–proton collisions.


In particle colliders, luminosity is a fundamental parameter that must be continuously monitored, as it directly determines the rate of particle interactions and, consequently, the statistical precision of physics measurements. It is defined as the proportionality factor between the interaction rate and the cross-section of a given process. Accurate knowledge of the luminosity is essential both for real-time monitoring of beam conditions and for the precise determination of cross-sections in physics analyses.

During Runs~3 and~4 of the LHC, the upgraded LHCb experiment operates at a luminosity approximately five times higher than in previous runs. This significant increase enhances the physics reach of the experiment while simultaneously imposing more stringent requirements on the performance and stability of the luminosity monitoring systems. Owing to its single-arm forward spectrometer design, the LHCb detector cannot operate at the same luminosity as the ATLAS and CMS experiments, since high occupancy can degrade the tracking performance in forward detectors. For this reason, the instantaneous luminosity is stabilised using a luminosity levelling technique~\cite{Muratori:1957033}. Consequently, real-time measurements of the instantaneous luminosity are essential for LHCb operations. The Probe for LUminosity MEasurement (PLUME), the luminometer of the upgraded LHCb detector~\cite{LHCb:2023hlw}, is designed to precisely monitor the LHC beam conditions at the LHCb interaction point. In particular, it provides real-time luminosity measurements to the levelling algorithm~\cite{Barsuk:2743098}, for which a precision of about 5\% is required. PLUME delivers online luminosity measurements with a statistical precision of about 0.3\% every 3 seconds, while the overall systematic uncertainty is of the order of 4\%, satisfying the requirements of the levelling system. In addition, it enables precise luminosity determination for offline analyses.

\begin{figure}[tbp]
\centering
\includegraphics[width=.8\textwidth,trim=30 110 0 0,clip]{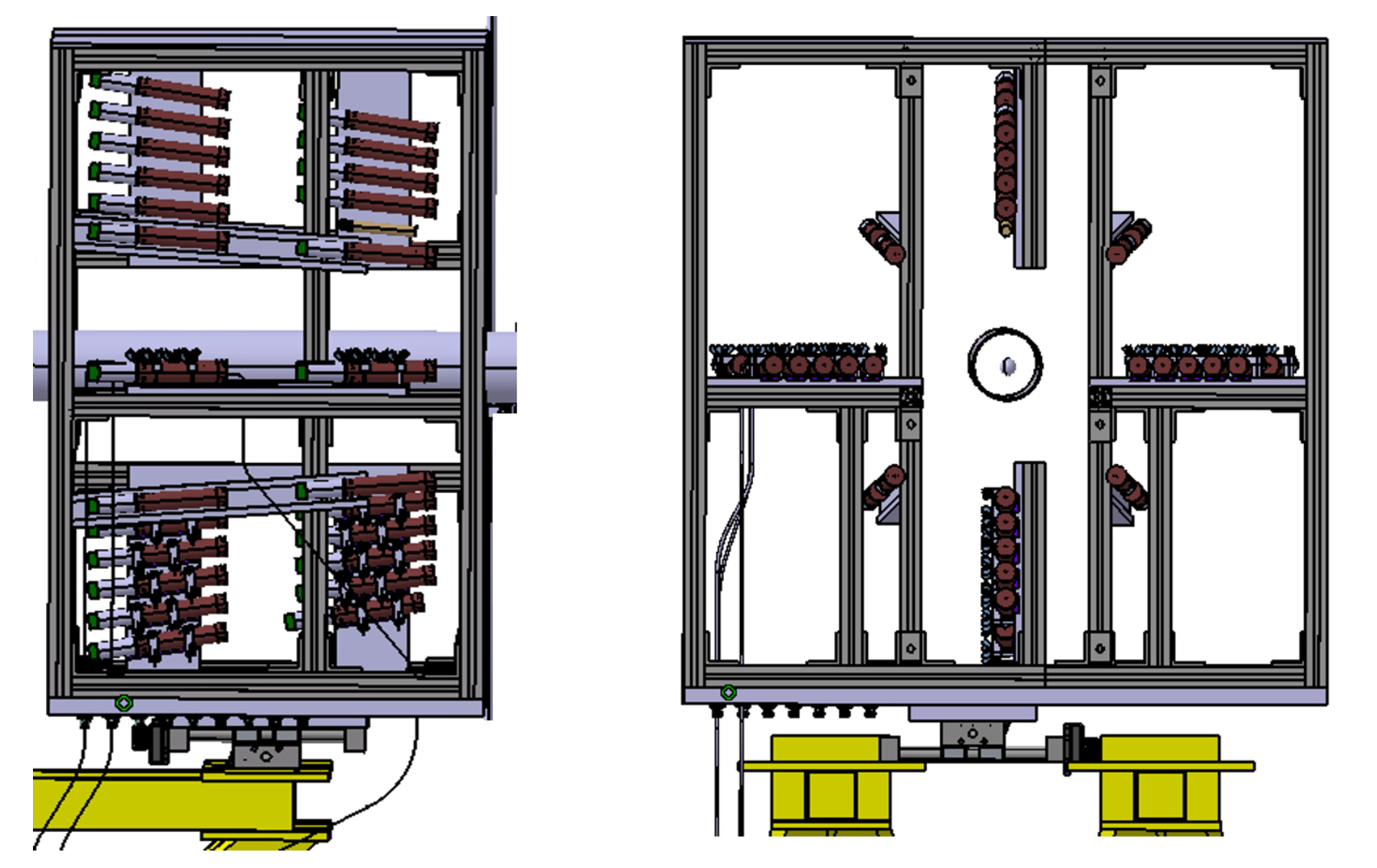}
\caption{\label{fig:PLUME_layout} Layout of the PLUME detector in the $y$--$z$ (left) and $x$--$y$ (right) planes, where $z$ is the beam axis. 
}
\end{figure}

\begin{figure}[htbp]
\centering 
\includegraphics[width=0.8\textwidth]{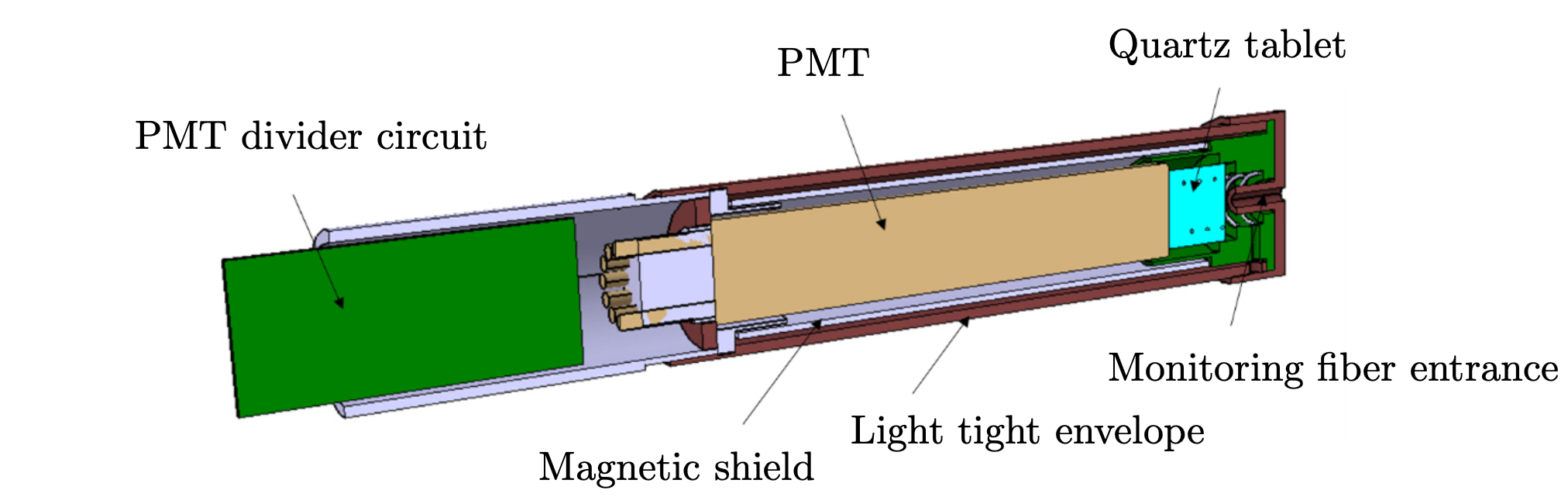}
\caption{\label{fig:detection_module} Schematic view of the PLUME elementary detection module. The module is 153 mm long with a diameter of 24 mm.}
\end{figure}

The PLUME detector is a hodoscope composed of 48 head-on-type Hamamatsu R760 photomultiplier tubes (PMTs), which provide luminosity measurements by detecting Cherenkov light produced in quartz by particles originating from the collision region. The detector extends from $z = -1485$ mm to $z = -2085$ mm along the beam axis and covers a pseudorapidity range of $-2.4 < \eta < -3.1$. The negative $z$ and $\eta$ values reflect the adopted coordinate convention, in which the positive $z$-axis is defined along the LHCb experiment direction, from the VELO towards the muon chambers; PLUME is therefore located in the opposite direction.
The PMTs are arranged in a projective geometry, forming a two-layer hodoscope with a cross-shaped layout around the beam pipe, as shown in Fig.~\ref{fig:PLUME_layout}. A 5~mm-thick quartz tablet is coupled to the PMT entrance window using optical grease, increasing the number of photons generated by incoming particles. Simulations indicate that this addition enhances the photon yield by approximately a factor of five. The PMT, quartz tablet, and divider circuit are housed within a cylindrical aluminium shield, forming the elementary detection module of the luminometer, as illustrated in Fig.~\ref{fig:detection_module}.

\begin{figure}[tbp]
\centering
\includegraphics[width=1\textwidth,trim=30 110 0 0,clip]{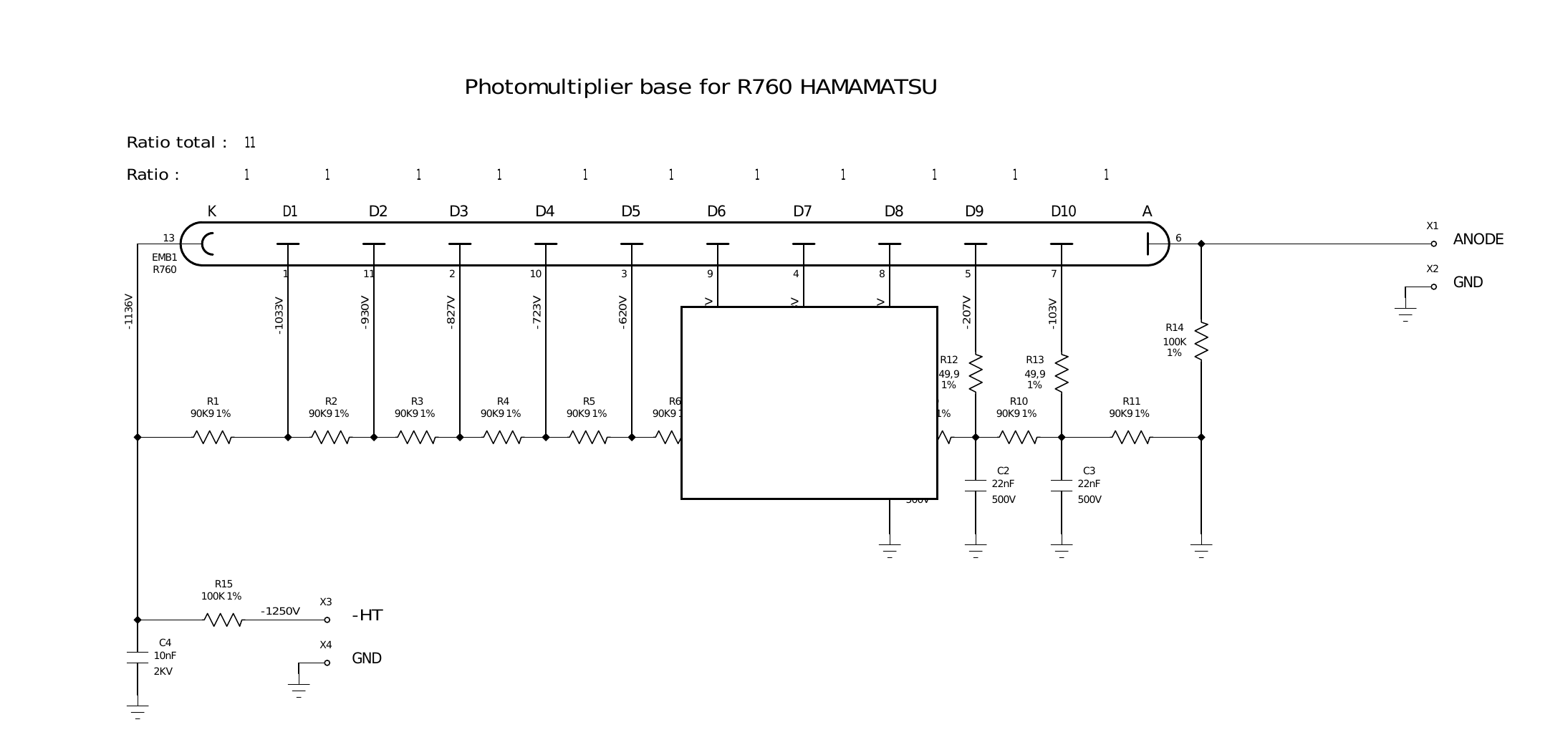}
\caption{\label{fig:divider} Schematic of the divider circuit used in the PLUME PMTs.}
\end{figure}

The PLUME detection unit consists of a head-on photomultiplier tube with ten linear-focused dynodes. The resistive voltage divider provided by Hamamatsu Photonics exhibits limited linearity for the signal amplitudes typically observed in PLUME, particularly in light of the increased photon yield introduced by the quartz tablet. For this reason, a custom divider circuit has been developed. It is designed to supply a maximum current of 1~mA at a bias voltage of 1000~V. A schematic of the custom divider, developed at IJCLab, is shown in Fig.~\ref{fig:divider}.

The harsh operating environment requires the use of radiation-resistant materials. Radiation tolerance is therefore a key requirement, as the expected dose in the relevant region ranges from 80 to 200~kGy, with a neutron fluence of approximately $1\times10^{14}$~n/cm$^2$. This integrated dose is expected to be accumulated by the detection modules by the end of Run~4. More detailed information about the PLUME detector can be found in the Technical Design Report~\cite{CERN-LHCC-2021-002}.

The performance of the detector depends critically on the characteristics of the PMTs. To ensure the correct operation of PLUME, the gain, transit-time drift, dark current, and linearity of all installed PMTs were measured prior to installation. In addition, an ageing campaign was conducted to study gain degradation as a function of the integrated charge during operation, and to assess whether the selected PMT model can withstand the full duration of Run~3 and Run~4 without replacement.

The paper is organized as follows. Following this introduction, Sec.~\ref{sec:exp_setup} describes the experimental setup used to measure the aforementioned characteristics of each detection module. The results of these measurements are presented in Sec.~\ref{sec:measurements}, and the conclusions are drawn in Sec.~\ref{sec:conclusions}.

\section{Experimental setup}
\label{sec:exp_setup}

The experimental setup used to characterise the PMTs varies slightly depending on the specific property under study. In all configurations, the device under test is mounted on a support and placed inside a light-tight metal enclosure to prevent external light from entering. The light source used to generate photoelectrons is a picosecond diode laser with driver from ALPHALAS, operating at a wavelength of $405$~nm and with a pulse full width at half maximum (FWHM) of $25$~ps~\cite{alphalas_picopower_ld}. The intensity of the light incident on the PMT is adjusted by inserting neutral density filters along the optical path. The laser can operate in both continuous-wave and pulsed modes. In pulsed mode, the repetition rate is set to $500$~Hz during data acquisition, due to limitations of the data acquisition board. During the ageing campaign, the light source was replaced with a LED from Thorlabs with a peak wavelength of $525$~nm, driven by an HP-8012B pulse generator. The PMTs are powered by a 40-channel CAEN SY127 system. The PMT output signal is acquired using different devices, depending on the measurement being performed:
\begin{itemize}
\item a $5^{1/2}$-digit Model 6485 picoammeter by Keithley, with a resolution of $1$ pA for currents below $200$ nA~\cite{keithley_6487_manual}. This instrument is used for relative gain and dark current measurements;
\item a DRS4 Evaluation Board provided by the Paul Scherrer Institute~\cite{psi_drs4_evalboard}. The DRS4 chip is a switched capacitor array capable of digitising eight channels at sampling speeds up to 5 GSPS with 1024 sampling points. This device is used to measure absolute gain, transit-time drift, linearity, and for the ageing campaign;
\item a Philips $2534$ system multimeter, used to monitor the PMT current and to measure the relative gain during the ageing campaign.
\end{itemize}
For the measurement of the PMT linearity response, the optical power of the laser is measured using a calibrated Thorlabs photodiode connected to a Thorlabs PM100 digital optical power meter~\cite{thorlabs_s122a}.

\section{Photomultiplier tube characterisation}
\label{sec:measurements}
In the following, the measurements performed during the characterisation campaign are described in detail, including those of gain, transit-time drift, linearity, dark current, and ageing.

\subsection{Gain}
\label{subsec:gain}

The detection modules register Cherenkov light emitted by particles produced in proton--proton collisions impinging on the PMT entrance window. Data collected during the commissioning phase of Run~3 indicate that, when operating at a gain of $1.5\times10^5$ and at the nominal luminosity of the LHCb experiment during Runs~3 and~4 (approximately $2000\ \mathrm{Hz}/\mu\mathrm{b}$), each PMT collects on average about $3\ \mathrm{pC}$ of integrated charge per particle. Under these conditions, the total charge accumulated by each PMT is expected to reach approximately 450~C over the course of Runs~3 and~4, as discussed in Sec.~\ref{sec:ageing}.
During operation, the PMT gain decreases due to the degradation of the dynode emissive materials and the photocathode layer. Measuring the gain as a function of the applied bias voltage for each PMT is therefore essential to determine the operating voltage required to maintain a stable gain.
Ideally, the gain can be determined from the ratio of the anode current to the photocathode current. However, since the latter is not directly accessible, an alternative approach is adopted based on measurements in the single-photoelectron (SPE) regime. In this configuration, each light pulse incident on the PMT window produces either zero or one photoelectron emitted from the photocathode. When a photoelectron is generated, the integrated charge measured at the anode corresponds to the PMT gain multiplied by the elementary charge.
The number of detected photoelectrons $k$ follows a Poisson distribution with mean $\nu$, such that the probability of observing $k$ photoelectrons is given by
\begin{equation}
\label{eq:poiss}
P(k) = \frac{\nu^k}{k!} e^{-\nu},
\end{equation}
where $\nu$ represents the average number of photoelectrons produced per light pulse. In terms of the number of recorded events, this can be written as $N_k = N_\mathrm{tot} P(k)$, where $N_\mathrm{tot}$ is the total number of events.

In the SPE regime, events with more than one photoelectron ($k>1$) are strongly suppressed and can be neglected. This condition is achieved by requiring the probability of zero-photoelectron events to be $P(k=0)=N_0/N_\mathrm{tot} \simeq 0.95$, corresponding to $\nu \simeq 0.05$ and implying a probability of multi-photoelectron events $P(k\geq 2) \simeq 10^{-3}$.
In practice, $\nu$ is estimated by measuring the fraction of events with signal amplitude above a fixed threshold, corresponding to events in which at least one photoelectron is produced, and the laser intensity is tuned accordingly to reach the SPE regime. After calibration, the noise of the DRS4 Evaluation Board is centred at zero with an RMS of $0.4$~mV. The amplitude threshold, denoted by $T$, is therefore set to $T \leq -3$~mV to ensure adequate separation from the baseline noise. Under these conditions, $\nu$ is evaluated as
\begin{equation}
\label{eq:nu_estimation}
\nu = -\ln(1 - R),
\end{equation}
where $R$ is the fraction of events with amplitude exceeding the threshold $T$. A value of $R = 0.05$ is chosen to ensure operation in the SPE regime.

\begin{figure}[!tbp]
\centering
\includegraphics[width=0.54\textwidth]{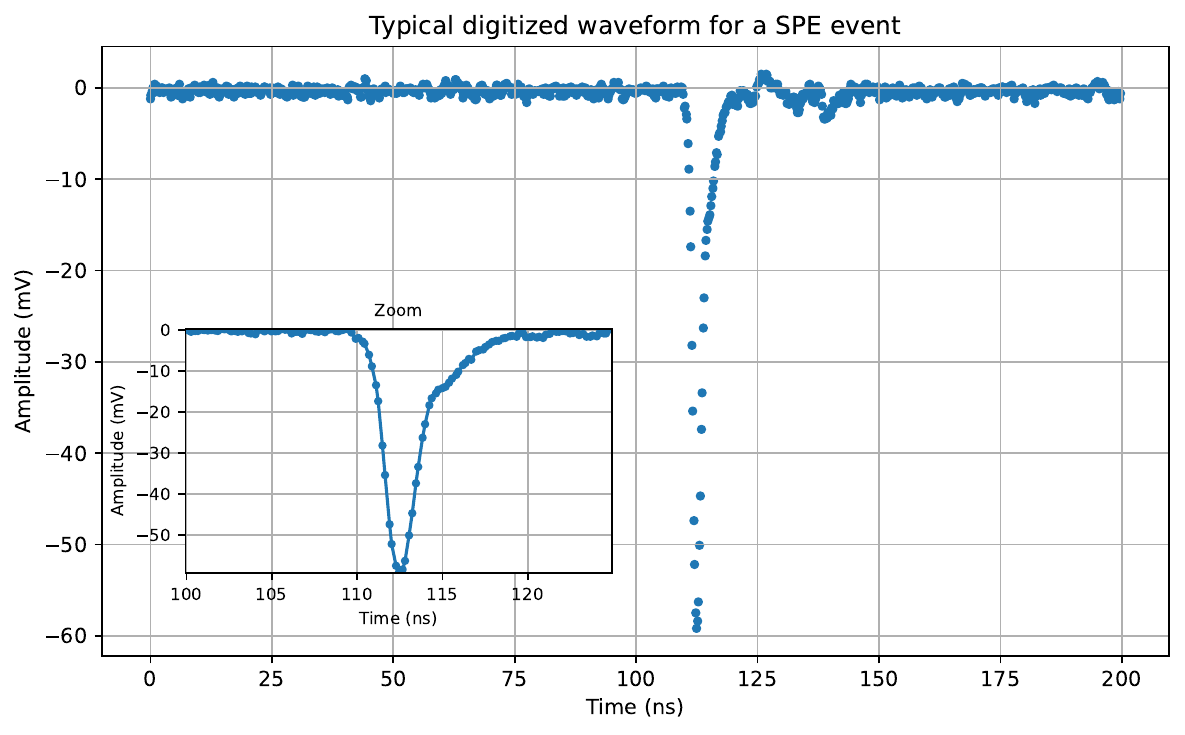}
\includegraphics[width=0.44\textwidth]{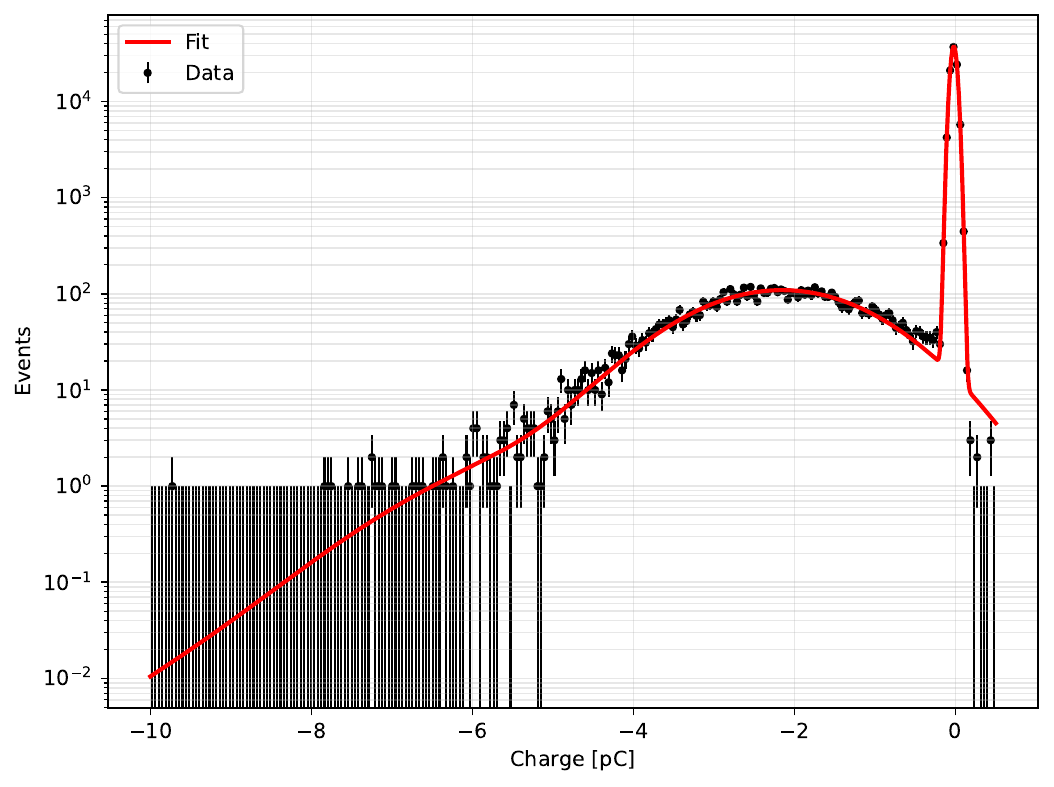}
\caption{\label{fig:SPE} Left: Typical digitized waveform for a single-photoelectron (SPE) event.
Right: Integrated charge distribution obtained from $10^5$ events recorded with PMT EA3544 at the maximum high voltage (1250 V) under SPE conditions. The black points represent the data, while the red line shows the fit result. The narrow peak around zero corresponds to pedestal events (no photoelectrons), while the broader peak centred around $-3$ pC corresponds to single-photoelectron events. The tail at lower charge values is due to events with multiple photoelectrons.}
\end{figure}

Once the SPE condition is achieved, the absolute gain is determined by fitting a model to the integrated charge spectrum obtained by integrating the digitized waveform of approximately $10^5$ events.  A typical SPE digitized waveform is shown in the left part of Fig.~\ref{fig:SPE}, while the integrated charge distribution obtained from $10^5$ events collected in the SPE regime is reported in the right part of Fig.~\ref{fig:SPE}. The narrow peak near zero represents the pedestal (electronics noise), corresponding to events without photoelectron emission. The broader contribution centered at around $-3$ pC corresponds to single photoelectron events. Data have been collected at the maximum allowed voltage (1250 V for this PMT model) to achieve a clear separation between pedestal and single photoelectron events.

\begin{figure}[tbp]
\centering
\includegraphics[width=0.49\textwidth]{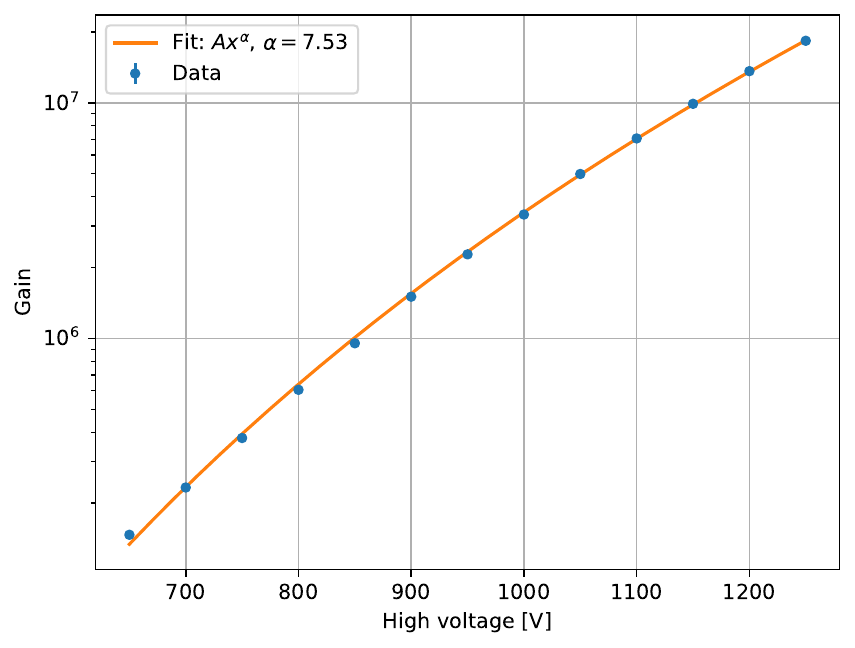}
\includegraphics[width=0.49\textwidth]{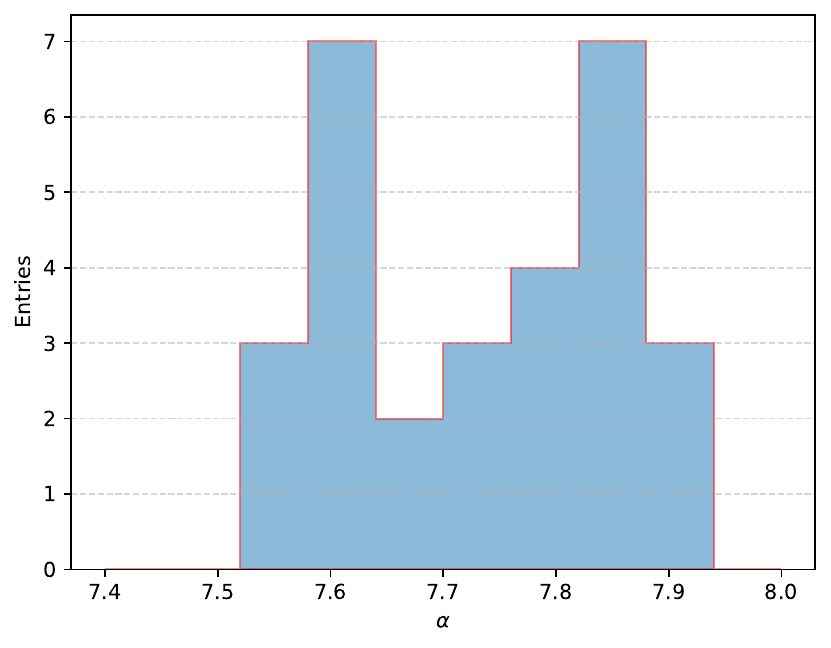}
\caption{\label{fig:gain_vs_hv} Left: Average gain as a function of the applied voltage for one of the tested PMTs, together with the corresponding fit.
Right: Distribution of the $\alpha$ values obtained from fits to the $G(V)$ measurements for all PMTs.}
\end{figure}

The integrated charge distribution is described by the convolution of Poisson and Gaussian functions~\cite{TAKAHASHI20181}
\begin{equation}
\label{eq:model fit}
f(x)= N\Biggl\{
e^{-\nu}\frac{1}{\sqrt{2\pi\sigma_\mathrm{p}^2}}
\exp\left(-\dfrac{(x-\mu_\mathrm{p})^2}{2\sigma_\mathrm{p}^2}\right)
+\sum_{k=1}^3\frac{\nu^k}{k!}e^{-\nu} \frac{1}{\sqrt{2\pi k\sigma_\mathrm{s}^2}}
\exp\left(-\dfrac{(x-k\mu_\mathrm{s})^2}{2k\sigma_\mathrm{s}^2}\right)
\Biggr\},
\end{equation}
where $N$ is a normalisation factor. The first term represents the pedestal contribution, characterised by a mean $\mu_\mathrm{p}$ and width $\sigma_\mathrm{p}$. The remaining terms describe events with $1 \leq k \leq 3$ photoelectrons, whose means and widths are given by $k\cdot\mu_\mathrm{s}$ and $\sqrt{k}\cdot\sigma_\mathrm{s}$, respectively.
The average gain is extracted using the relation
\begin{equation}
\label{eq:gain}
G_{\mathrm{max}} = \frac{ \hat{\mu}_\mathrm{s} - \hat{\mu}_\mathrm{p}}{q_e},
\end{equation}
where $\hat{\mu}_\mathrm{s}$ and $\hat{\mu}_\mathrm{p}$ are the best-fit values of the single-photoelectron peak and pedestal means, respectively, and $q_e$ is the elementary charge.
Once $G_{\mathrm{max}}$ is determined, the gain as a function of the applied bias voltage is studied. The laser is operated in continuous-wave mode, providing a constant light intensity, and the anode is measured with the picoammeter at 13 voltage values ranging from $650$~V to $1250$~V in steps of $50$~V. The gain at a given voltage is then obtained as
\begin{equation}
\label{eq:gain_proportion}
G(V) = G_{\mathrm{max}} \frac{I(V)}{I_{\mathrm{max}}},
\end{equation}
where $I(V)$ is the measured anode current at voltage $V$, and $I_{\mathrm{max}}$ is the current measured at the maximum voltage.
Assuming equal potential differences between the dynodes, the gain can be expressed as
\begin{equation}
\label{eq:power_law}
G(V) = \delta^n = C \cdot (\Delta V^\rho)^n,
\end{equation}
where $\delta$ is the gain per dynode, $n$ is the number of dynodes, $\Delta V$ is the voltage difference between successive dynodes, $\rho$ is a parameter that depends on the dynode material and structure, and $C$ is a proportionality constant. Defining $\alpha \equiv \rho\cdot n$, Eq.~\eqref{eq:power_law} can be rewritten as
\begin{equation}
\label{eq:power_law2}
G(V) = C \cdot \Delta V^\alpha.
\end{equation}
For the Hamamatsu R760 model, $\alpha$ is expected to lie in the range $7$--$8$, since $n=10$ and $\rho \approx 0.7$--$0.8$~\cite{r760hamamatsu,handbook}. The value of $\alpha$ is determined by fitting Eq.~\eqref{eq:power_law2} to the measured $G(V)$ values for each PMT. The gain as a function of the applied voltage for a representative PMT is shown in the left panel of Fig.~\ref{fig:gain_vs_hv}, together with the result of the fit. The same figure also shows the distribution of the $\alpha$ values obtained for all tested PMTs, which is consistent with expectations.

\subsection{Transit-time drift}

Knowledge of the timing characteristics of the detection modules is crucial to ensure that the signals produced by particles in all PLUME PMTs are fully contained within the 25 ns LHC bunch-crossing interval, thereby preventing spillover into subsequent bunch crossings. The Hamamatsu R760 model features a transit time of approximately $22$~ns and a rise time of about $2.1$~ns~\cite{r760hamamatsu}. The time response of the PMT signal is primarily determined by the electron transit time, $t_\mathrm{transit}$, defined as the time required for photoelectrons emitted at the photocathode to reach the anode.

The transit time is measured as the time interval between the laser trigger signal and the appearance of the PMT output pulse. In practice, both the laser trigger and the PMT output signals are acquired using a digitiser, and their timestamps are determined using a constant-fraction method at $50\%$ of the signal amplitude. The timestamps $t_1$ (laser trigger) and $t_2$ (PMT signal) are obtained through linear interpolation between the samples immediately before and after the $50\%$ threshold crossing. All measurements are performed while maintaining the PMT output integrated charge at approximately $6$~pC, ensuring a consistent charge density at the last dynode stages.
The transit time is then defined as
\begin{equation}
\label{eq:ttd}
t_\mathrm{transit}(V) = t_2 - t_1,
\end{equation}
where the dependence on the applied voltage $V$ has been made explicit.

\begin{figure}[!htbp]
\centering
\includegraphics[width=0.6\textwidth]{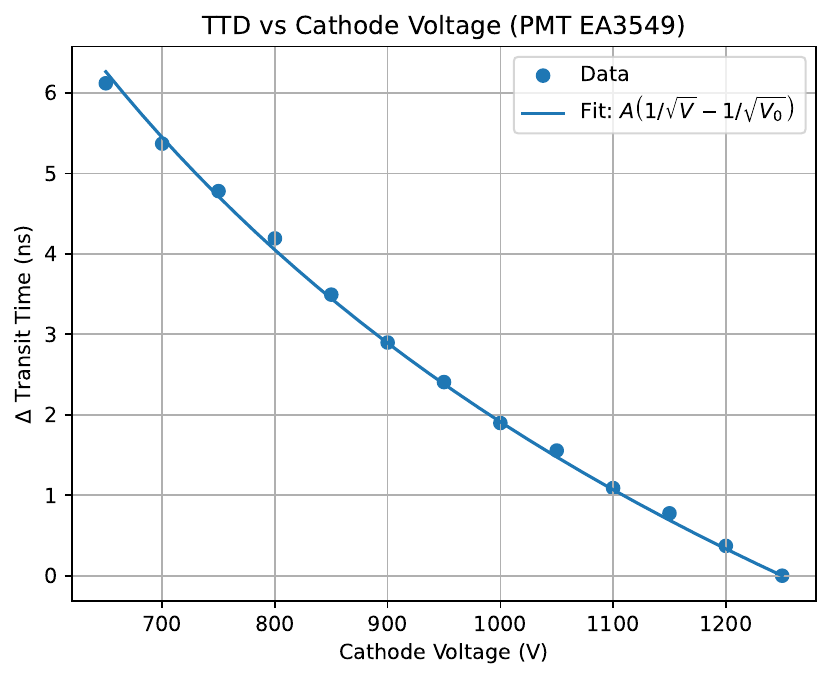}
\caption{\label{fig:TTD_EA3564} Transit-time drift as a function of the applied supply voltage for PMT EA3564, computed with respect to the value measured at 1250~V.}
\end{figure}

To study the dependence of $t_\mathrm{transit}$ on the applied voltage, measurements of the transit-time drift, $\Delta t_\mathrm{transit}$, are performed. The transit-time drift is defined as the variation of the transit time with respect to a reference value measured at a fixed voltage. It is computed as
\begin{equation}
\label{eq:relative_ttd}
\Delta t_\mathrm{transit}(V) = t_\mathrm{transit}(V) - t_\mathrm{transit}(1250~\mathrm{V}),
\end{equation}
where $t_\mathrm{transit}(1250~\mathrm{V})$ is the transit time measured at the maximum operating voltage, used as the reference point. Figure~\ref{fig:TTD_EA3564} shows the transit-time drift for a representative PMT as a function of the applied voltage. A fit with a function proportional to $V^{-1/2}$, reflecting the expected dependence of the transit time on the applied voltage, is also shown and is found to describe the data well. The maximum value of $\Delta t_\mathrm{transit}$ does not exceed 7~ns for any of the tested PMTs. Since the R760 signal width is approximately 10~ns and the LHC bunch-crossing interval is 25~ns, even a drift of this magnitude has a limited impact on the PLUME performance, provided that the signals are well centred within the bunch-crossing time window.

\subsection{Linearity}

The linearity of a PMT characterises its ability to produce an output signal, either in terms of anode current or integrated charge, that is proportional to the number of photoelectrons generated at the photocathode, and hence to the intensity of the incident light, within a given dynamic range. In the PLUME detector, the modules are exposed to pulsed light signals produced by particles originating from the LHCb experiment interaction point. Under these conditions, deviations from linearity arise primarily from space-charge effects in the last dynode stages and at the anode, which limit the charge amplification at high signal intensities. The luminosity measurement algorithm is based on the logZero method~\cite{LHCb:2014set}, which exploits the Poisson statistics of bunch crossings with no signal above threshold. While this method is in principle robust against non-linear effects, they can still alter the detection efficiency of individual modules and introduce biases in the measured luminosity. It is therefore essential to characterise the linearity of each detection module and to operate the PMTs within their linear response regime.

\begin{figure}[tbp]
\centering
\includegraphics[width=0.7\textwidth]{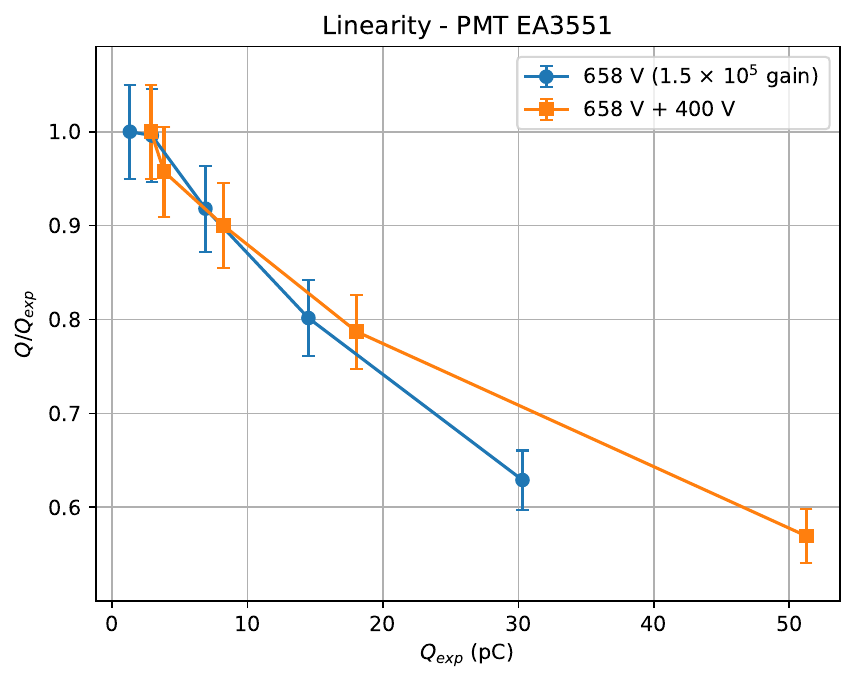}
\caption{\label{fig:linearity_EA3551} Measured-to-expected charge ratio as a function of the expected charge for PMT EA3551. The blue (orange) points and lines correspond to measurements performed at the operational voltage (400 V above the operational voltage). A relative uncertainty of 5\% is assigned to each data point, reflecting the limited precision of the low-light-intensity measurements associated with the SPE regime.}
\end{figure}

The linearity is studied at two different supply voltages: the voltage corresponding to a gain of $1.5\times10^5$ (approximately 650~V) for each PMT, and a voltage 400~V higher, representative of the conditions expected at the end of Run~4. For each voltage, measurements are performed at five different light intensities. The attenuation at each step is chosen such that the light intensity is approximately halved with respect to the previous setting. The integrated charge collected by the PMT is measured for each configuration by recording $10^4$ events, from which the average integrated charge, $Q$, is computed. The expected integrated charge, $Q_\mathrm{exp}$, is obtained by rescaling the charge measured in SPE regime, denoted as $Q_\mathrm{ref}$, by the ratio of the light intensity at the given operating point, $P$, to the reference light intensity, $P_\mathrm{ref}$:
\begin{equation}
\label{eq:exp_charge}
Q_\mathrm{exp} = Q_\mathrm{ref}\frac{P}{P_\mathrm{ref}}.
\end{equation}
The reference point is chosen since the PMTs are expected to behave linearly when, on average, only a single photoelectron is produced. Due to the limited precision in measuring the low light intensities associated with the SPE regime, a relative uncertainty of $5\%$ is assigned to each data point. In the case of perfect linearity, the ratio $Q/Q_\mathrm{exp}$ is expected to be equal to unity. As shown in Fig.~\ref{fig:linearity_EA3551}, deviations of up to $30\%$ are observed for integrated charges of about 40~pC. However, linearity is preserved within 10\% for integrated charges below 10~pC, corresponding to the upper limit of the PLUME operational regime. On average, the observed deviation from linearity across all PMTs at 10~pC translates into a bias of about 6\% in the luminosity measurement. Since the average integrated charge per bunch crossing during operation is closer to 3~pC, the corresponding average luminosity bias is reduced to about 1.5\%. Within the associated uncertainties, this result remains compatible with the absence of any significant bias in the luminosity determination arising from non-linearity effects. The results obtained at the two high-voltage settings are compatible within uncertainties, indicating that the linearity does not strongly depend on the applied gain.

\subsection{Dark current}
\begin{figure}[tbp]
\centering
\includegraphics[width=0.7\textwidth]{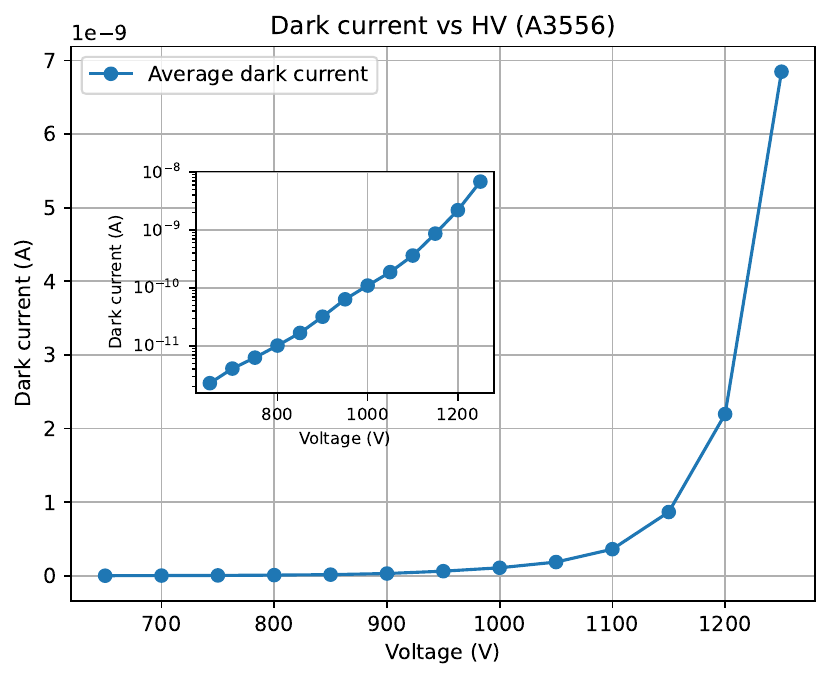}
\caption{\label{fig:darkCurrent_EA3556} Dark current as a function of the applied voltage for PMT EA3556. The inset shows the same data on a semi-logarithmic scale.}
\end{figure}

The dark current is the small current that flows in a photomultiplier tube even when operating in complete darkness, and it is primarily due to the spontaneous emission of electrons from the photocathode. The dark current depends on the applied high voltage, although this dependence is not linear. Measuring this quantity is important, as large dark current values could lead to spurious signals, thereby biasing the luminosity measurement.
The dark current is measured for all PMTs at 13 fixed values of the supply voltage, ranging from 650~V to 1250~V in steps of 50~V. The PMTs are kept under high voltage in the dark box for 30 minutes prior to the measurement to ensure stable operating conditions. For each voltage setting, twenty current measurements are recorded at a rate of one per second and averaged to obtain the final value. As expected, the dark current increases with increasing voltage. However, it remains below 10~nA even at the maximum supply voltage, as shown in Fig.~\ref{fig:darkCurrent_EA3556} for PMT EA3556. The highest dark current measured among all tested PMTs is approximately 20~nA. Since the average anode current during nominal operation is of the order of several $\mu$A, the contribution of the dark current can be considered negligible.

\subsection{Ageing}
\label{sec:ageing}

The PMTs installed in the PLUME detector are required to operate throughout LHC Runs 3 and 4 without replacement. When a module is exposed to high particle rates over extended periods, several of its operational characteristics may change. This gradual degradation in performance is referred to as ageing. In general, ageing depends on factors such as the operating history, the secondary emissive material of the dynodes, and the output current level, and it typically results in a decrease in both sensitivity and gain.

Assuming a collision rate of 30 MHz, a geometrical acceptance of 6\%, an average of 125 photoelectrons per detected particle, 180 days of operation per year, and a gain of $1.5 \times 10^5$, each PMT is expected to accumulate an integrated charge of approximately 80 C per year. Since Runs 3 and 4 are expected to span about 5.5 years of data taking, the PMTs should therefore withstand a total integrated charge of roughly 450 C. It is therefore essential to verify that gain degradation due to ageing can be compensated by increasing the applied voltage, while remaining below the maximum operating voltage of the R760 model.

\begin{figure}[!tbp]
\centering
\includegraphics[width=0.49\textwidth]{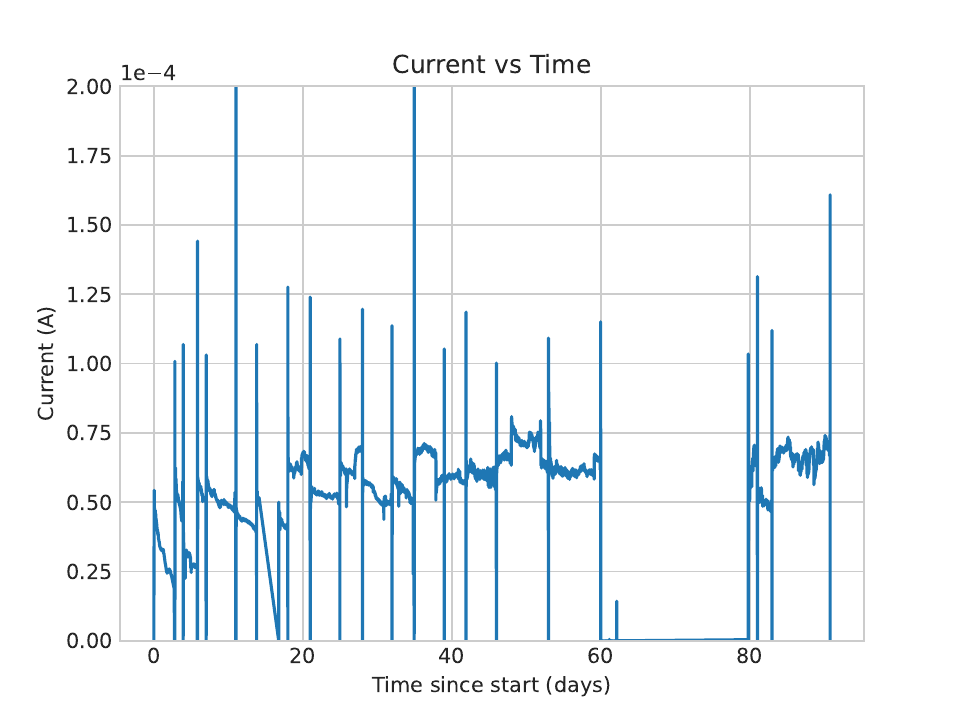}
\includegraphics[width=0.49\textwidth]{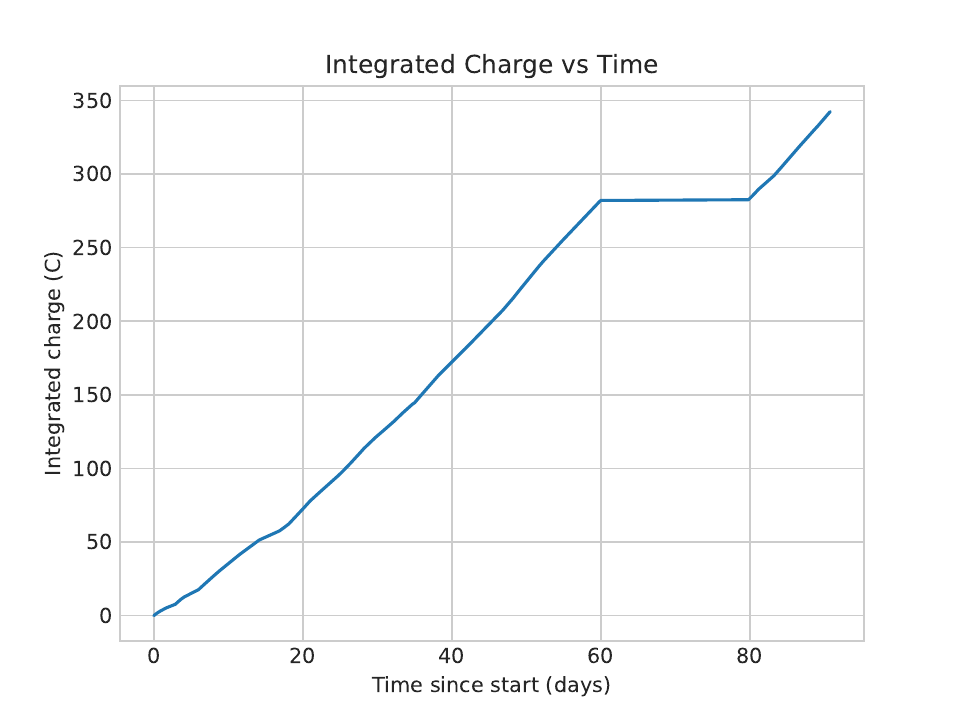}
\caption{\label{fig:ageing_campaign} Left: Anode current measured by the multimeter during the ageing campaign. The peaks, reaching values of about $100\ \mu\mathrm{A}$, correspond to periods during which the absolute and relative gain were measured.
Right: Integrated charge accumulated during the ageing campaign. The plateau between the 60th and 80th day corresponds to a pause in the ageing process.}
\end{figure}

An ageing campaign was carried out by illuminating a PMT placed inside a dark box with 525 nm LED light. During ageing, the LED was operated in a quasi-continuous regime, with a pulse frequency of 50 MHz and a pulse width of 20 ns. At regular intervals of a few days, the pulse frequency was reduced to 500 Hz (due to limitations of the DRS4 data acquisition rate), and the LED intensity was adjusted to reach the SPE regime in order to measure the absolute gain of the PMT. After this measurement, the LED was returned to quasi-continuous operation, and the average anode current was measured at seven different voltage values to determine the gain relative to the absolute one.
During the ageing process, the anode current was continuously monitored every ten seconds using a multimeter connected to a computer for readout, as shown in Fig.~\ref{fig:ageing_campaign}. The vertical spikes, reaching values of about $100\ \mu\mathrm{A}$, correspond to interruptions of the ageing procedure required for gain measurements. After each measurement, the current was not restored to the same value due to difficulties in operating the analogue waveform generator used during the campaign.

The campaign was divided into three phases, as shown in the right panel of Fig.~\ref{fig:ageing_campaign}. After an initial ageing period of approximately 60 days, during which the PMT accumulated about 280 C of integrated charge, the process was halted for 20 days to investigate possible gain recovery during inactivity. Subsequently, the ageing process was resumed up to a total integrated charge of about 340 C, exceeding the value expected to be accumulated during Run 3 and approaching the total integrated charge anticipated by the end of Run 4. 

\begin{figure}[!tbp]
\centering
\includegraphics[width=0.49\textwidth]{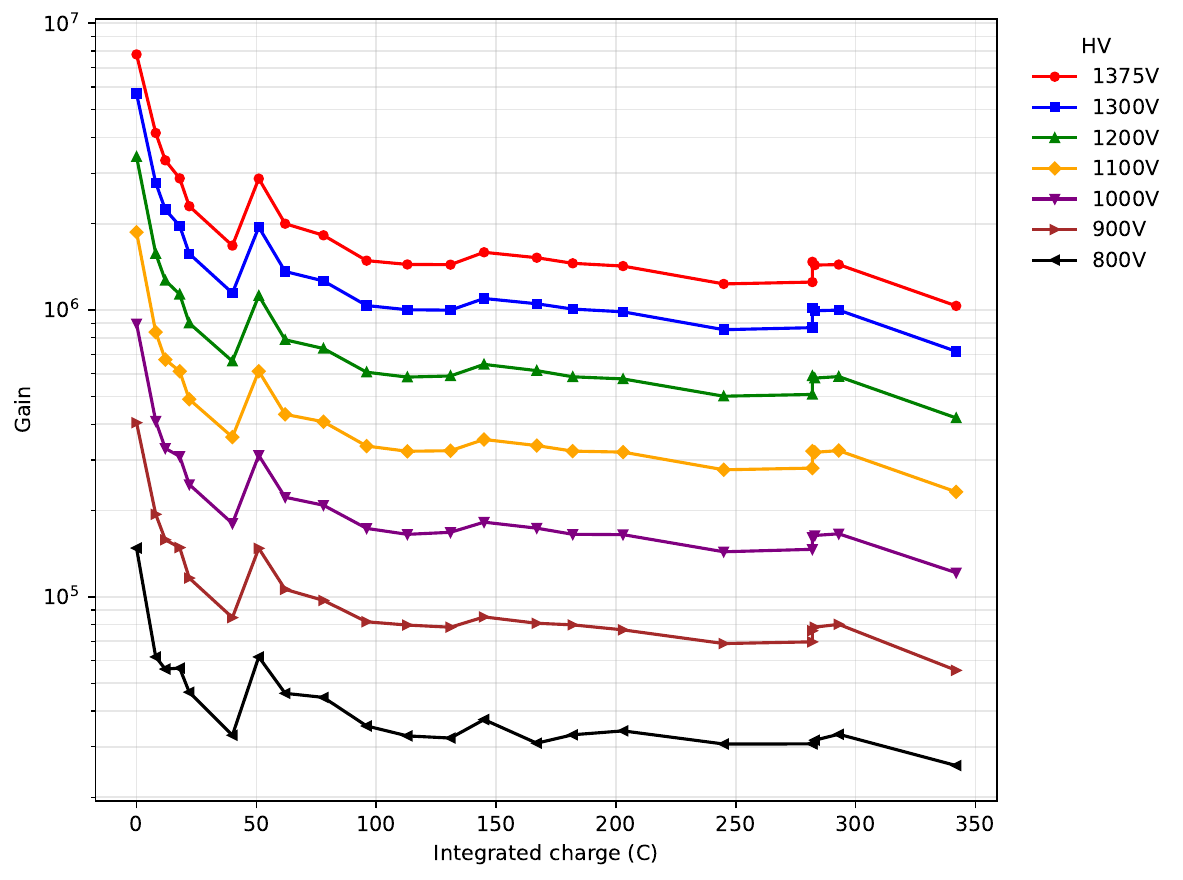}
\includegraphics[width=0.49\textwidth]{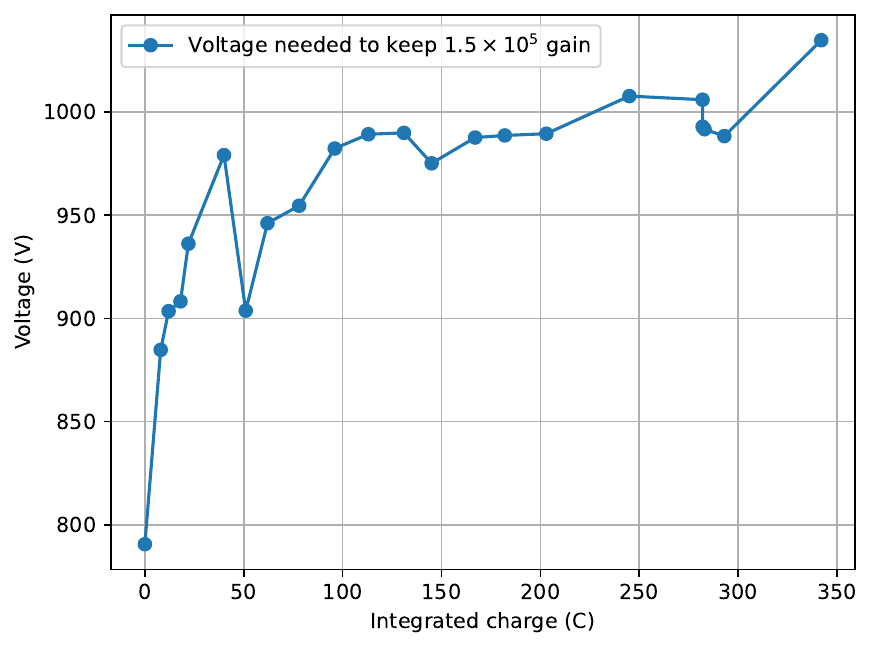}
\caption{\label{fig:ageing_gain} Left: Gain as a function of the integrated charge collected by the PMT for several high-voltage values.
Right: Voltage applied to the PMT to maintain a gain of $1.5 \times 10^5$ as a function of the integrated charge.}
\end{figure}

As expected, the gain decreases with increasing integrated charge, as shown in the left panel of Fig.~\ref{fig:ageing_gain} for several high-voltage values. After a rapid initial decrease by about a factor of four within the first 40 C, the gain exhibits a slower, approximately linear decline. Using the relative gain measurements, it is possible to determine the increase in high voltage required to compensate for the gain loss. The voltage needed to maintain a gain of $1.5 \times 10^5$ is shown in the right panel of Fig.~\ref{fig:ageing_gain} as a function of the integrated charge.

The pause in the ageing process starting after the 60th day does not show any significant recovery of the PMT gain. Indeed, any gain recovered during the period of inactivity is quickly lost once charge accumulation resumes. At the end of the campaign, the high voltage required to maintain a constant gain of $1.5 \times 10^5$ reached approximately 1035 V, corresponding to an increase of about 250 V relative to the initial value, and remained well below the maximum allowed value for the Hamamatsu R760. These results indicate that the PLUME detection modules are capable of operating throughout LHC Runs 3 and 4 without requiring replacement.

\section{Conclusions}
\label{sec:conclusions}

The PLUME detector is a hodoscope composed of 24 pairs of PMTs arranged in a projective geometry, designed to provide both online and offline luminosity measurements for the LHCb experiment. The selected PMT model is the Hamamatsu R760, chosen for its fast signal response, radiation hardness, and its successful use in the LUCID luminometer of the ATLAS experiment during LHC Runs 1 and 2~\cite{Avoni_2018}. The measurement of the main PMT characteristics constitutes a fundamental part of the quality assurance process. In this work, we have presented a summary of the measurements performed by INFN and IJCLab prior to the installation of the PMTs in PLUME. The gain, transit-time drift, linearity, and dark current of each PMT were measured and found to be within the expected ranges.

The average absolute gain at the maximum supply voltage of 1250 V is found to be $1.5\times 10^7$. 
The supply voltage required to achieve a gain of $1.5\times10^5$ was also determined for each PMT. These values define the operational settings for the initial data-taking phase of Run 3. The dependence of the transit time on the supply voltage was quantified through measurements of the transit-time drift. The variation between the minimum and maximum operating voltages was found to be smaller than 7 ns, indicating a limited sensitivity of the transit time to changes in the supply voltage. Furthermore, since the Hamamatsu R760 signal has a typical duration of approximately 10 ns and the LHC bunch-crossing interval is 25 ns, even the maximum observed transit-time shift remains compatible with containing the signal within a single bunch crossing. The linearity of the PMT response was tested at two different supply voltages. Although deviations from linearity of up to 10\% are observed at the upper end of the PLUME operational range (10 pC), the detector typically operates at average integrated charges of about 3 pC per bunch crossing. At this level, the corresponding bias on the luminosity measurement is estimated to be only about 1.5\% and remains compatible with zero within the associated uncertainties. The PMT response can therefore be regarded as effectively linear for the luminosity measurements performed by PLUME. The dark current of all PMTs remains below 20 nA even at the highest operating voltage and can be considered negligible compared to the signal current expected during Runs 3 and 4.
An ageing campaign was performed to verify that the PMTs can withstand the full duration of LHC Runs 3 and 4. During this campaign, a total integrated charge of 340 C was accumulated, slightly lower than the amount expected by the end of Run 4. After an initial gain reduction of approximately a factor of four, the gain exhibited a slow and steady decrease. To maintain a constant gain of $1.5 \times 10^5$, the high voltage had to be increased from about 800 V to 1035 V. This value remains well below the maximum allowable voltage for the R760 model. These results demonstrate that, with appropriate adjustment of the high voltage, the Hamamatsu R760 PMTs will be capable of operating throughout the entire duration of Runs 3 and 4 without requiring replacement.

\acknowledgments
Luca Toscano appreciates support by the DFG within the Emmy Noether Program under grant number MI 2869/1-1.


 \bibliographystyle{JHEP}
 \bibliography{biblio.bib}

@article{Muratori:1957033,
      author        = "Muratori, B. and Pieloni, T.",
      title         = "{Luminosity levelling techniques for the LHC}",
      archivePrefix = "arXiv",
      eprint        = "1410.5646",
      pages         = "177-181",
      year          = "2014",
      url           = "https://cds.cern.ch/record/1957033",
      note          = "Comments: 5 pages, contribution to the ICFA Mini-Workshop
                       on Beam-Beam Effects in Hadron Colliders, CERN, Geneva,
                       Switzerland, 18-22 Mar 2013",
      doi           = "10.5170/CERN-2014-004.177",
}

@techreport{CERN-LHCC-2021-002,
      collaboration = "LHCb",
      title         = "{LHCb PLUME: Probe for LUminosity MEasurement}",
      institution   = "CERN",
      reportNumber  = "CERN-LHCC-2021-002, LHCB-TDR-022",
      address       = "Geneva",
      year          = "2021",
      url           = "https://cds.cern.ch/record/2750034",
      doi           = "10.17181/CERN.WLU0.M37F",
}

@article{TAKAHASHI20181,
title = {A technique for estimating the absolute gain of a photomultiplier tube},
journal = {Nuclear Instruments and Methods in Physics Research Section A: Accelerators, Spectrometers, Detectors and Associated Equipment},
volume = {894},
pages = {1-7},
year = {2018},
issn = {0168-9002},
doi = {https://doi.org/10.1016/j.nima.2018.03.034},
url = {https://www.sciencedirect.com/science/article/pii/S016890021830370X},
author = {M. Takahashi and Y. Inome and S. Yoshii and A. Bamba and S. Gunji and D. Hadasch and M. Hayashida and H. Katagiri and Y. Konno and H. Kubo and J. Kushida and D. Nakajima and T. Nakamori and T. Nagayoshi and K. Nishijima and S. Nozaki and D. Mazin and S. Mashuda and R. Mirzoyan and H. Ohoka and R. Orito and T. Saito and S. Sakurai and J. Takeda and M. Teshima and Y. Terada and F. Tokanai and T. Yamamoto and T. Yoshida},
}

@misc{r760hamamatsu,
  author       = {{Hamamatsu Photonics}},
  title        = {R760 Photomultiplier Tube Data Sheet},
  howpublished = {\url{https://www.hamamatsu.com/us/en/product/optical-sensors/pmt/pmt_tube-alone/head-on-type/R760.html}},
  note         = {Accessed: 2026-03-27}
}

@misc{handbook,
  author       = {{Hamamatsu Photonics}},
  title        = {Photomultiplier Tubes: Basics and Applications},
  howpublished = {\url{https://www.hamamatsu.com.cn/content/dam/hamamatsu-photonics/sites/documents/99_SALES_LIBRARY/etd/PMT_handbook_v4E.pdf}},
  note         = {Accessed: 2026-03-27}
}

@misc{psi_drs4_evalboard,
  author       = {{Paul Scherrer Institute (PSI)}},
  title        = {{DRS4} Evaluation Board},
  howpublished = {\url{https://www.psi.ch/en/ltp-muon-physics/evaluation-board}},
  note         = {Accessed: 2026-03-30}
}

@manual{keithley_6487_manual,
  author       = {{Keithley Instruments, Inc.}},
  title        = {Model 6487 Picoammeter/Voltage Source User's Manual},
  year         = {2011},
  howpublished = {\url{https://download.tek.com/manual/6487-900-01(C-Mar2011)(User).pdf}},
  note         = {Revision C, Accessed: 2026-03-30}
}

@manual{alphalas_picopower_ld,
  author       = {{ALPHALAS GmbH}},
  title        = {Picosecond Pulse Diode Lasers with Driver: PICOPOWER-LD Series Datasheet},
  year         = {},
  howpublished = {\url{https://www.alphalas.com/images/stories/products/lasers/Picosecond_Pulse_Diode_Lasers_with_Driver_PICOPOWER-LD_ALPHALAS.pdf}},
  note         = {Accessed: 2026-03-30}
}

@manual{thorlabs_s122a,
  author       = {{Thorlabs, Inc.}},
  title        = {S120/S122/S210/S212 Optical Power Meter System Manual},
  howpublished = {\url{https://media.thorlabs.com/globalassets/items/s/s1/s12/s122a/12271-d02.pdf}},
  note         = {Includes S122A photodiode sensor specifications, Accessed: 2026-03-30}
}

@techreport{Barsuk:2743098,
      author        = "Barsuk, Sergey and Roy, Laurent and Rochet, Jacky and
                       Panshin, Gennady and Sanders, Freek and Balagura, Vladislav
                       and Fleuret, Frederic and Maurice, Emilie Amandine and
                       Beigbeder-Beau, Christophe and Burmistrov, Leonid and
                       Cornebise, Patrick and Machefert, Frederic and Niel,
                       Elisabeth Maria and Puill, Veronique and Robbe, Patrick and
                       Carbone, Angelo and Galli, Domenico and Marconi, Umberto
                       and Perazzini, Stefano and Vagnoni, Vincenzo and Cholak,
                       Serhii and Graverini, Elena and Haefeli, Guido and
                       Shchutska, Lesya and Van Dijk, Maarten and Guz, Iouri and
                       Orlov, Vladyslav and Yeroshenko, Vsevolod and
                       Konoplyannikov, Anatoli and Bezshyyko, Oleg and
                       Golinka-Bezshyyko, Larysa and Brault, Sylvian and Chaumat,
                       Vincent and Boyarintsev, Andrii and Ferrari, Fabio and
                       Nguyen Trung, Thi",
      title         = "{Probe for LUminosity MEasurement in LHCb}",
      institution   = "CERN",
      reportNumber  = "LHCb-PUB-2020-008, CERN-LHCb-PUB-2020-008",
      address       = "Geneva",
      year          = "2020",
      url           = "https://cds.cern.ch/record/2743098",
}

@article{Avoni_2018,
doi = {10.1088/1748-0221/13/07/P07017},
url = {https://doi.org/10.1088/1748-0221/13/07/P07017},
year = {2018},
month = {jul},
publisher = {},
volume = {13},
number = {07},
pages = {P07017},
author = {Avoni, G. and Bruschi, M. and Cabras, G. and Caforio, D. and Dehghanian, N. and Floderus, A. and Giacobbe, B. and Giannuzzi, F. and Giorgi, F. and Grafström, P. and Hedberg, V. and Manghi, F. Lasagni and Meneghini, S. and Pinfold, J. and Richards, E. and Sbarra, C. and Cesari, N. Semprini and Sbrizzi, A. and Soluk, R. and Ucchielli, G. and Valentinetti, S. and Viazlo, O. and Villa, M. and Vittori, C. and Vuillermet, R. and Zoccoli, A.},
title = {The new {LUCID-2} detector for luminosity measurement and monitoring in {ATLAS}},
journal = {Journal of Instrumentation}
}

@article{LHCb:2023hlw,
    author = "Aaij, Roel and others",
    collaboration = "LHCb",
    title = "{The LHCb Upgrade I}",
    eprint = "2305.10515",
    archivePrefix = "arXiv",
    primaryClass = "hep-ex",
    reportNumber = "LHCb-DP-2022-002",
    doi = "10.1088/1748-0221/19/05/P05065",
    journal = "JINST",
    volume = "19",
    number = "05",
    pages = "P05065",
    year = "2024"
}

@article{LHCb:2014set,
  author         = {Aaij, Roel and others},
  collaboration  = {LHCb},
  title          = {Precision luminosity measurements at {LHCb}},
  journal        = {JINST},
  volume         = {9},
  year           = {2014},
  pages          = {P12005},
  doi            = {10.1088/1748-0221/9/12/P12005},
  eprint         = {1410.0149},
  archivePrefix  = {arXiv},
  primaryClass   = {hep-ex}
}


%
%
%
%
\end{document}